\documentclass[twocolumn,
 amsmath, amssymb, breqn,
 aps, prl, superscriptaddress, twocolumn,
longbibliography]{revtex4-1}

\usepackage{graphicx}% Include figure files
\usepackage{dcolumn}% Align table columns on decimal point
\usepackage{bm}% bold math
\usepackage{upgreek}
\usepackage{tikz}
\usepackage{wasysym}

\begin{document}

\renewcommand{\b}[1]{{\boldsymbol{#1}}}

\title{Stabilized Pair Density Wave via Nanoscale Confinement of Superfluid $^3$He}

\author{A.J. Shook}
 \affiliation{Department of Physics, University of Alberta, Edmonton, Alberta T6G 2E1, Canada} 
 
\author{V. Vadakkumbatt}%
 \affiliation{Department of Physics, University of Alberta, Edmonton, Alberta T6G 2E1, Canada} 

\author{P. Senarath Yapa}%
 \affiliation{Department of Physics, University of Alberta, Edmonton, Alberta T6G 2E1, Canada} 

\author{C. Doolin}%
 \affiliation{Department of Physics, University of Alberta, Edmonton, Alberta T6G 2E1, Canada} 

\author{R. Boyack}%
 \affiliation{Department of Physics, University of Alberta, Edmonton, Alberta T6G 2E1, Canada} 
 \affiliation{Theoretical Physics Institute, University of Alberta, Edmonton, Alberta T6G 2E1, Canada}
 
 \author{P.H. Kim}%
 \affiliation{Department of Physics, University of Alberta, Edmonton, Alberta T6G 2E1, Canada} 
 
 \author{G.G. Popowich}%
 \affiliation{Department of Physics, University of Alberta, Edmonton, Alberta T6G 2E1, Canada} 
 
 \author{F. Souris}%
 \affiliation{Department of Physics, University of Alberta, Edmonton, Alberta T6G 2E1, Canada} 

 \author{H. Christani}%
 \affiliation{Department of Physics, University of Alberta, Edmonton, Alberta T6G 2E1, Canada} 
   
\author{J. Maciejko}
\email{maciejko@ualberta.ca}
 \affiliation{Department of Physics, University of Alberta, Edmonton, Alberta T6G 2E1, Canada} 
 \affiliation{Theoretical Physics Institute, University of Alberta, Edmonton, Alberta T6G 2E1, Canada}
 
\author{J.P. Davis}
\email{jdavis@ualberta.ca}
 \affiliation{Department of Physics, University of Alberta, Edmonton, Alberta T6G 2E1, Canada} 

\date{\today}

\begin{abstract}

Superfluid $^3$He under nanoscale confinement has generated significant interest due to the rich spectrum of phases with complex order parameters that may be stabilized.  Experiments have uncovered a variety of interesting phenomena, but a complete picture of superfluid $^3$He under confinement has remained elusive.  Here, we present phase diagrams of superfluid $^3$He under varying degrees of uniaxial confinement, over a wide range of pressures, which elucidate the progressive stability of both the $A$-phase, as well as a growing region of stable pair density wave (PDW) state.  
\end{abstract}

\pacs{Valid PACS appear here}% PACS, the Physics and Astronomy
                             % Classification Scheme.
%Use showkeys class option if keyword
                              %display desired
\maketitle

\addtolength{\textfloatsep}{-0.1in}

While bulk superfluid $^3$He is exceptionally well understood, both theoretically and experimentally \cite{Vollhardt13}, much is unknown when confinement approaches the scale of the superfluid coherence length.  In the bulk, only two phases are stable, the so-called $A$ and $B$ phases.  The $A$-phase---with nodes in the gap, as shown in Fig.~1(a)---is stabilized by strong coupling effects at high pressures \cite{Rainer76}, whereas the $B$-phase---with its isotropic gap---dominates the phase diagram (Fig.~4(a)).  When confined via engineered structures, the phase diagram is expected to be altered significantly.  Due to the unconventional $p$-wave pairing in $^3$He, non-magnetic scattering at surfaces is sufficient to break Cooper pairs.  Thus surface scattering serves to distort and suppress the order parameter.  In fact, under confinement the $B$-phase is altered to become the planar-distorted $B$-phase, see Eq.~(\ref{pdBphase}), with a suppressed gap at antipodal points on the Fermi surface (Fig.~1(a)).  Early experiments to explore the effects of confinement resulted in a variety of observations, including hints of a new phase transition \cite{Xu90}, the complete elimination of the $B$-phase in favor of the $A$-phase \cite{Freeman90}, and modification of the relative stability of the $A$ and $B$-phases \cite{Miyawaki00,Kawasaki04}.

It was realized by Vorontsov and Sauls that this order parameter suppression due to surface scattering could be minimized by the formation of domain walls between two orientations of planar-distorted $B$-phases, Fig.~1(b), when the confinement lies in a particular range of length scales \cite{Vorontsov05}. Alignment of these domains walls was later predicted to form the basis of an ordered superfluid phase, deemed the stripe phase \cite{Vorontsov07}.  The stripe phase is an example of a pair density wave (PDW), a state that breaks both gauge and translational symmetries and is now believed to play an important role in the cuprate superconductors \cite{Agterberg19}.  The observation of such a PDW state in superfluid $^3$He could lead to a deeper understanding of this state, and influence understanding of PDWs in unconventional superconductors \cite{Holmvall18}.

\begin{figure}[b]
    \includegraphics[width=.45\textwidth]{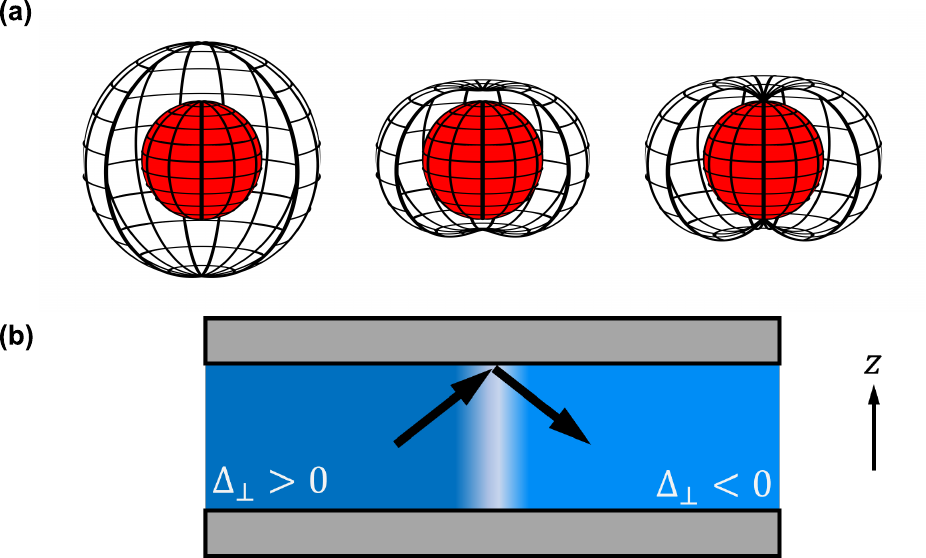}
    \caption{(a) Schematic momentum-space representations of the quasiparticle gaps for the relevant superfluid states under planar confinement: from left to right, the isotropic $B$-phase, the planar distorted $B$-phase, and the $A$-phase.  (b) Confinement along the $z$-axis. When Cooper pairs scatter off a wall the $z$-component $\Delta_\perp$ of the order parameter changes sign, which for the planar-distorted $B$-phase causes pair breaking.  This can be minimized by the formation of domain walls between regions of planar-distorted $B$-phase with alternating signs of $\Delta_\perp$. A regular arrangement of these domain walls leads to broken translational symmetry and the PDW state.} 
    \label{fig:phases}
    \centering
\end{figure}

The prediction of Vorontsov and Sauls stimulated numerous experimental searches, including nuclear magnetic resonance (NMR) \cite{Levitin13,Levitin19}, shear micromechanical resonators \cite{Zheng16}, and torsional oscillators \cite{Zhelev17}---with spatial confinements of 700 nm \cite{Levitin13}, 1.1 $\mu$m \cite{Levitin19,Zhelev17}, and 1.25 $\mu$m \cite{Zheng16}. Most conclusively, NMR studies have simultaneously observed signals from two different planar-distorted $B$-phases.  Despite this observation, they have concluded that their measurements are inconsistent with the stripe phase, and instead have suggested another PDW with soft domain walls between localized puddles of planar-distorted $B$-phase, which they have termed the polka-dot phase \cite{Levitin19}.  To date, it is unclear if these results represent the true thermodynamic phase diagram of superfluid $^3$He under uniform confinement, or if they remain complicated by non-uniformities in the structures, e.g., bowing induced by pressurization.  Here, in an effort to resolve this question, we use microfabricated fourth-sound resonators \cite{Rojas15,Souris17}, which are sensitive indicators of the superfluid fraction, to explore the generation of spatial order in superfluid $^3$He.  These devices are fully immersed in liquid and therefore are mechanically unaffected by pressurization of the $^3$He.  This allows us to explore the phase diagram up to 28 bar, revealing signatures of PDW in the phase diagram that have never been observed.  Furthermore, we simultaneously measure three devices with varying degrees of confinement, in order to map the effect of confinement on the stability of the PDW state.  

\begin{figure}[t]
    \includegraphics[width=.45\textwidth]{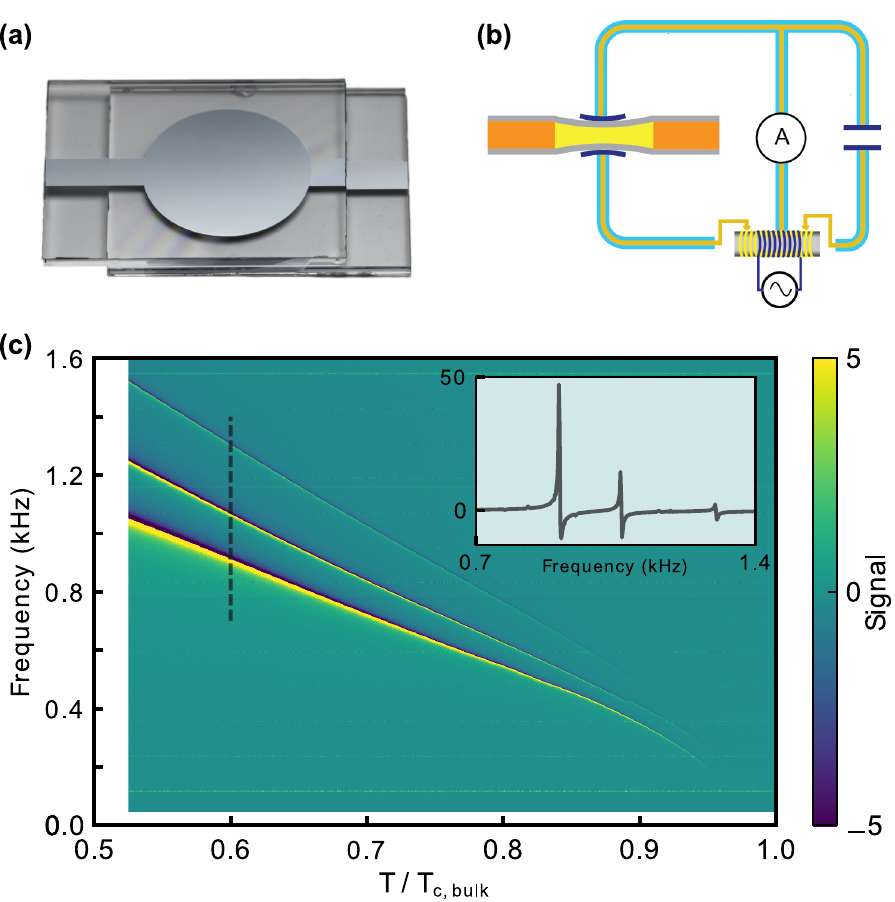}
    \caption{Experimental system and superfluid Helmholtz resonator frequency data. (a) Helmholtz resonator, fabricated out of single-crystal quartz etched to define the nanoscale confinement, then deposited with aluminum inside the etched area.  (b) Under an applied voltage the electrodes are slightly displaced towards each other, pushing the superfluid $^3$He out of the channel and exciting the Helmholtz mode.  This is detected using a capacitance bridge that outputs a current that is digitized.  (c) Three Helmholtz resonators, with varying confinement, are wired in parallel and measured simultaneously using a chirped pulse scheme described in the text, while warming the sample cell via remagnetization of the copper nuclear stage \cite{SM}.  Due to the measurement configuration the resonances appear Fano-like, yet are easily resolved, as shown in the inset---corresponding to the cross-section marked with the dashed line in the main panel. }
    \label{fig:Device}
    \centering
\end{figure}

The devices used to probe the superfluid phases under confinement, shown in Fig.~2(a), are microfabricated from single-crystal quartz, with similar devices having been previously demonstrated to selectively probe the superfluid fraction within a nanoscale channel of $^4$He \cite{Rojas15}.  They work by exciting a mechanical resonance in the fluid upon excitation by a voltage source, Fig.~2(b).  This causes a deflection of the two electrodes---on opposite sides of the nanoscale cavity---via a strong electrostatic force due to the small gap between the electrodes.  As a result of the deflection, fluid in the nanoscale channels is pushed out, which---when driven on resonance---excites a Helmholtz mechanical mode.  In the devices measured here, with channel depths of $636\pm12$ nm, $805\pm4$ nm, and $1067\pm7$ nm, the normal fluid is viscously clamped \cite{Kojima75}, hence the frequency $\omega$ of the Helmholtz mode is related to the superfluid fraction $\rho_s/\rho$:
\begin{equation}
\label{Eq:resonance}
\omega^2=K\left(\frac{\rho_s}{\rho}\right).
\end{equation}
$K$ represents terms that involve measured geometric factors, and the calibrated spring constant of the plate and helium \cite{Rojas15}.  In this work, we have extended this to superfluid $^3$He by constructing an adiabatic nuclear demagnetization stage and $^3$He sample cell \cite{Pollanen12,Krusius78}, described further in the Supplemental Material \cite{SM}. 

\begin{figure}[b]
    \includegraphics[width=.45\textwidth]{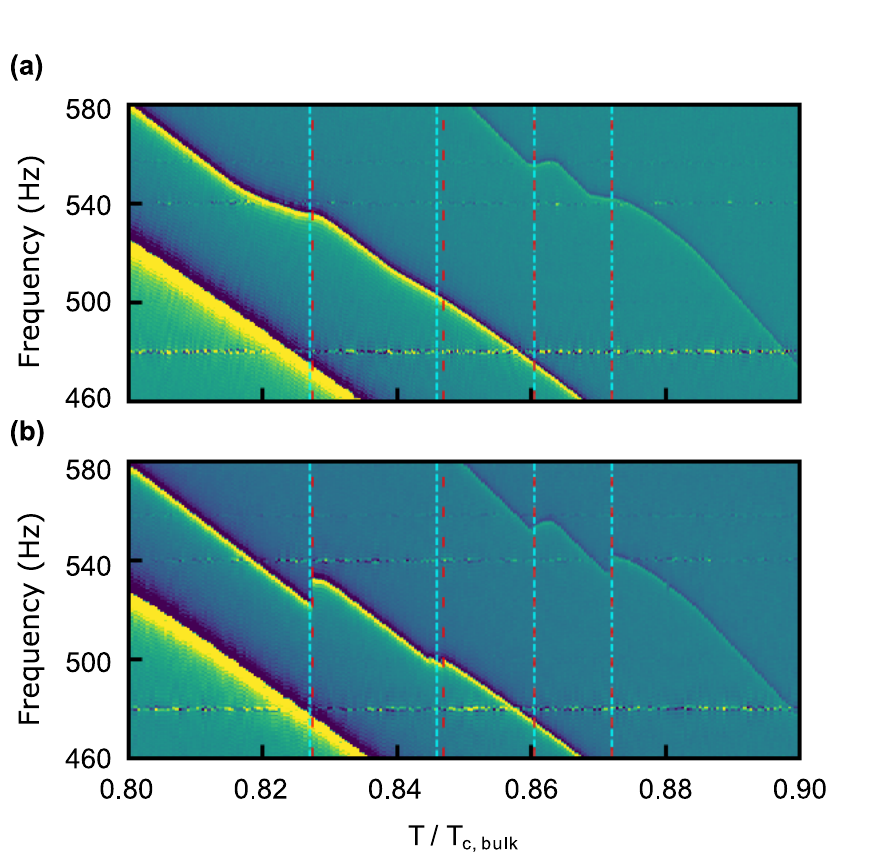}
    \caption{Zoom-in around phase transitions for 22 bar data.  (a) Cooling run with phase transitions in the 1067 nm device (upper trace) and 805 nm device (middle trace) marked with blue dashed lines.  (b) Same as in (a) but for a warming run and phase transitions marked with red dashed lines.  Note that (a) and (b) are on the same scales and the dashed lines cross both panels to emphasize that there is no significant temperature hysteresis between warming and cooling in our experiments.  The hysteresis in shape of the transitions can be understood from the evolution of latent heat at first order phase transitions, as observed in the shape hysteresis between the A and B-phases in bulk superfluid $^3$He \cite{Osheroff72,Richardson}.  Observation of two first-order phase transitions demonstrates not only that the PDW state at intermediate temperatures is stable \cite{Wiman16}, but the lack of hysteresis points to an intermediate phase between the $A$ and planar distorted $B$-phases that eliminates supercooling \cite{Zhelev17,Tye11}. }
    \label{fig:transitions}
    \centering
\end{figure}

\begin{figure*}[t]
    \includegraphics[width=.95\textwidth]{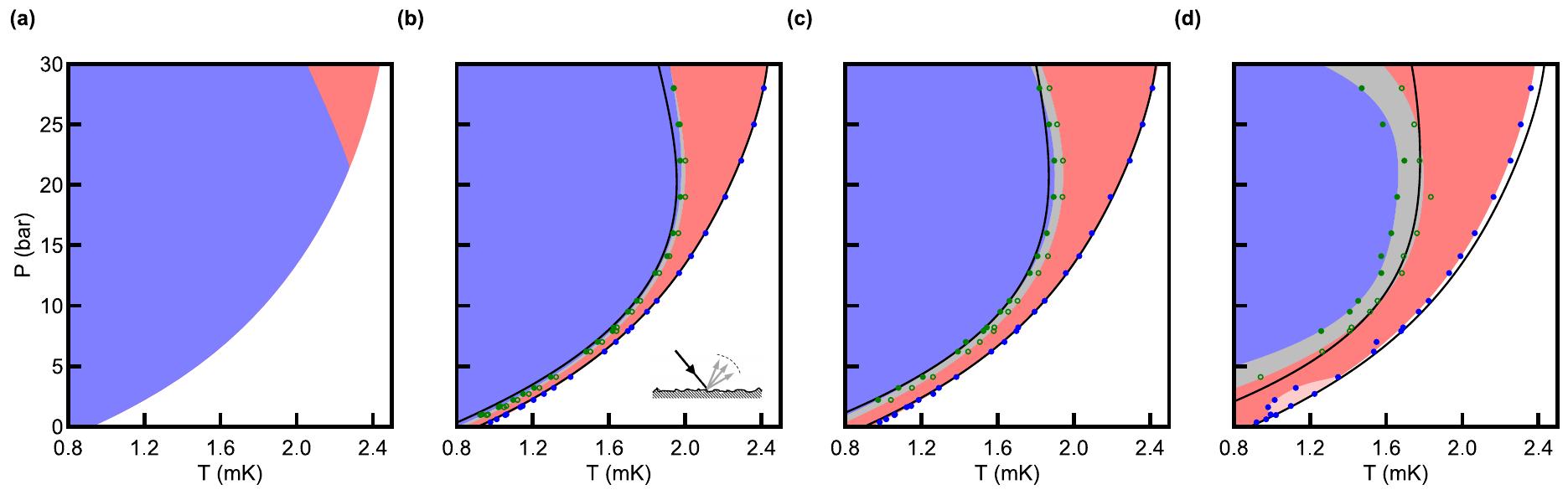}
    \caption{Pressure ($P$)-temperature ($T$) phase diagrams of superfluid $^3$He, on warming.  (a) Bulk phase diagram shown for reference, with the $A$-phase (pink) and $B$-phase (blue).  Phase diagrams for (b) 1067 nm, (c) 805 nm, and (d) 636 nm confinements are shown with $A$-phase (pink) and planar-distorted $B$-phase (blue).  Under confinement, a new phase appears, which we have colored grey.  As the confinement increases from left to right, the width of the grey region grows, as does the stability of the $A$-phase.  Ginzburg-Landau calculations, including strong coupling corrections \cite{Choi07,Wiman16} and the effect of confinement \cite{Ho1988} with diffuse boundary conditions (illustrated schematically in the inset of (b)), are included as the black curves---without any adjustable parameters. The PDW must lie between the $A$-phase and the planar distorted $B$-phase  \cite{Vorontsov07,Wiman16}, hence the grey region is concluded to be the stable PDW state. }
 \label{fig:phasediagrams}
 \centering
\end{figure*}

To drive the superfluid mechanical mode, we use a chirped pulse scheme to excite all three devices simultaneously and acquire their response in the time domain \cite{Doolin19}.  Fast Fourier transforming this time domain data allows us to rapidly acquire the frequency response of the Helmholtz resonators, and enables mapping of their frequencies during adiabatic nuclear demagnetization temperature sweeps.  An example at 12.7 bar is shown in Fig.~2(c).  Three superfluid-mechanical resonances are seen, with Fano-like character due to the measurement scheme, as shown in the inset.  The highest frequency resonance corresponds to the 1067 nm device and is assumed to have a temperature dependence that is indiscernible from that of the bulk superfluid fraction \cite{Parpia85,Wu13}, as was found in Ref.~\cite{Zhelev17}.  We use the temperature dependence of the 1067 nm device as a sensitive local secondary thermometer, which is referenced to both a melting curve thermometer on the same experimental stage and a tuning fork \cite{Carless83,Blaauwgeers07} immersed in the same sample cell.  We note that we have performed sweeps of the drive amplitude, and are in a linear response regime in all measurements.  This also serves to verify that our drive does not cause heating of the liquid $^3$He.  Thermal effects are discussed further in the Supplemental Material \cite{SM}.

The temperature dependence of the superfluid fraction is known to be an excellent indicator of phase transitions \cite{Zhelev17}.  We show in Fig.~3 portions of temperature sweeps, at 22 bar, demonstrating two phase transitions per device---suggesting the existence of three stable superfluid phases.  This observation is quite startling, not only as there are only two phases observed in bulk $^3$He (in zero magnetic field), but also that two distinct first-order phase transitions were not observed in previous NMR or torsional oscillator experiments.  Instead, those experiments found a single first-order phase transition that occurred over a broad temperature range \cite{Zhelev17}, which they attributed to inhomogeneous thickness of the superfluid from bowed devices.  Recent NMR experiments at low pressure in a 1.1 $\mu$m device do not show a broad phase transition, yet also do not observe two first-order phase transitions \cite{Levitin19}.  This is reasonable, considering the complete phase diagram that we show next.

To determine the phase diagram under confinement, and the relation between our data and previous observations, we have explored these phase transitions over a wide pressure range, from 0.35 bar to 28 bar, and compiled the thermodynamic warming transitions in Fig.~4. Note, we discuss the lack of temperature hysteresis between warming and cooling below.  In panels (b--d) of Fig.~4, the experimentally determined phase transitions are shown as circles: blue for the transition to the normal state, and the open and filled green circles for kinks in the superfluid fraction.  Smooth fits to the data points delineate colored regions, and hence regions of stable phases. 

To understand these results, we theoretically model the system using a Ginzburg-Landau approach. Superfluid $^3$He is a spin-triplet $p$-wave superfluid characterized by the Cooper pair amplitude
\begin{align}
\Delta_{\alpha\beta}(\hat{\b{p}})=A_{\mu j}(i\sigma_\mu\sigma_y)_{\alpha\beta}\hat{p}_j
\end{align}
between fermions of spin $\alpha,\beta$ and relative momentum $\hat{\b{p}}$, where $\sigma_x,\sigma_y,\sigma_z$ are the Pauli spin matrices and $A_{\mu j}$ is a $3\times 3$ matrix order parameter. We numerically solve the Ginzburg-Landau equations for $A_{\mu j}$ in a slab geometry following the formalism of Refs.~\cite{Wiman15,Wiman16} which incorporates confinement effects~\cite{Ho1988} as well as strong coupling corrections~\cite{Choi07}. The depairing effect of surface roughness~\cite{Ambegaokar74} on the cavity walls is taken into account through diffuse scattering boundary conditions. The order parameter profile is assumed to be $z$-dependent to account for confinement effects, but translationally invariant in the plane of the device ($x$-$y$ plane). Two such uniform phases are known to be stabilized under confinement, the planar-distorted $B$-phase with order parameter
\begin{align}\label{pdBphase}
    A_{\mu j}^\text{pdB}(z)=\left(\begin{array}{ccc}
    \Delta_\parallel(z) & 0 & 0 \\
    0 & \Delta_\parallel(z) & 0 \\
    0 & 0 & \Delta_\perp(z)
    \end{array}\right)_{\mu j},
\end{align}
and the $A$-phase, with order parameter
\begin{align}\label{Aphase}
    A_{\mu j}^\text{A}(z)=\left(\begin{array}{ccc}
    0 & 0 & 0 \\
    0 & 0 & 0 \\
    \Delta(z) & i\Delta(z) & 0 
    \end{array}\right)_{\mu j}.
\end{align}
Strong coupling corrections are necessary to account for the stability of the $A$-phase relative to the planar phase, i.e., Eq.~(\ref{pdBphase}) with vanishing $z$-component of the order parameter, $\Delta_\perp=0$.

We include the known bulk phase diagram in Fig.~4(a) for comparison, with the $B$-phase in blue and the $A$-phase in pink. The leftmost black curve in panels (b--d) is the calculated phase boundary between the planar-distorted $B$-phase and the $A$-phase, while the black curve on the right is the calculated pressure-dependent critical temperature $T_c(P)$. The excellent agreement between theory and experiment in panels (b--c) and to a lesser extent (d), as well as comparison with the bulk phase diagram, unambiguously suggest the blue and pink regions in the phase diagrams under confinement are the planar-distorted $B$-phase and $A$-phase, respectively. However, under the assumption of translational invariance in the $x$-$y$ plane, the Ginzburg-Landau analysis does not capture the experimentally observed grey region that lies between the two uniform phases.

In light of recent experimental and theoretical work, a clear candidate for this intermediate region is a spatially inhomogeneous phase exhibiting domain walls across which the $z$-component $\Delta_\perp$ of the superfluid order parameter changes sign, see Fig.~1(b). While in bulk $^3$He such domain walls~\cite{Salomaa88,Silveri14} are not energetically favorable, for device thicknesses $D$ of order the coherence length $\xi$, the energy cost of creating a domain wall can be compensated by the reduction in gradient energy associated with surface pairbreaking. Everywhere on the domain wall, $\Delta_\perp$ vanishes due to the change of sign and is thus uniform along the confinement direction, which reduces this energy. For sufficiently thin devices the gain in energy from the reduction in surface pairbreaking outweighs the cost from domain wall surface tension, and domain walls are favored. The enhanced stability of the grey region with increasing degree of confinement, observed in Fig.~4(b--d) and predicted in Ref.~\citenum{Wiman16}, thus further supports the interpretation of this region as a spatially inhomogeneous phase. 

It is noteworthy that such a spatially inhomogeneous phase can only exist between the $A$ and $B$-phases, as discussed in Ref.~\cite{Wiman16}, strengthening our assignment of the intermediate phase as a PDW state.  The stabilization of domain walls in the vicinity of the $A$-$B$ phase boundary can be understood from the following generic argument. Upon approaching the $A$-$B$ phase boundary from the planar-distorted $B$-phase, $\Delta_\perp$ is gradually reduced. Deep in the planar-distorted $B$-phase, suppressing $\Delta_\perp$ on a domain wall is not energetically favorable, but becomes increasingly so upon approach to the transition. In the $A$-phase, $\Delta_\perp=A_{zz}$ is uniformly zero due to the different symmetry of the phase, see Eq.~(\ref{Aphase}), thus surface pairbreaking is automatically minimized without the need for domain walls. Near the $A$-$B$ phase boundary, the free energies of the $A$-phase (with $\Delta_\perp=0$) and $B$-phase (with $\Delta_\perp\neq 0$) are nearly degenerate, and an inhomogeneous state with a regular arrangement of domain walls is a natural way to resolve the competition between the two different bulk ordering tendencies for $\Delta_\perp$.

While this argument alone cannot predict the optimal spatial arrangement of domain walls, previous studies strongly suggest periodic patterns of ordering are favored. Explicit calculations~\cite{Vorontsov07,Wiman16,Aoyama16} using either quasiclassical Green's functions or the Ginzburg-Landau approach, and for a variety of boundary conditions ranging from specular to maximally pairbreaking, indicate a periodic arrangement of domain walls forming a unidirectional PDW phase (stripe phase) can be stabilized at low pressures under confinement near the $A$-$B$ phase boundary for slab thicknesses $D\sim 10\xi$. However, NMR studies~\cite{Levitin19} favor an interpretation in terms of a two-dimensional PDW phase---the polka-dot phase---with the symmetry of a square or triangular lattice. In the pressure range 0--15 bar the PDW phase appears in our devices in a temperature range such that $8\lesssim D/\xi\lesssim 18$, in broad agreement with theory expectations for the stripe phase, but unexpectedly persists at pressures up to 30 bar, at least in the two thinnest devices.

Our current experiments cannot resolve the stripe/polka-dot debate, but do indicate that the PDW state %likely
observed here is thermodynamically stable from the fact that it is seen reproducibly both on warming and on cooling, as shown for example in Fig.~3.  It is also noteworthy that we observe no significant hysteresis in the temperatures of phase transitions between the $A$-phase and the PDW state, nor between the PDW state and the planar-distorted $B$-phase, despite the fact that these are first-order phase transitions for which supercooling would be expected---especially as significant supercooling between the $A$ and $B$-phases is routinely observed in bulk $^3$He.  The lack of hysteresis is consistent with recent torsional oscillator experiments \cite{Zhelev17}.  In that work, the existence of an unseen spatially modulated phase was suggested to account for the lack of hysteresis.  This can be understood by the presence of an intermediate phase between the $A$ and $B$-phases (the PDW) lowering the nucleation barrier between these states \cite{Tye11}. 

Finally, we note that we have observed an anomalous region in the 636 nm device between pressures of 0.95--4.10 bar, as seen in the light pink region of Fig.~4(d).  Whereas the rest of our phase transitions are fully reproducible, in this region we find that on some runs we observe transitions to the $A$-phase as expected, yet on occasion we find that the transition occurs at a reduced temperature.  Further experiments are required to elucidate the behavior of this region and will be the subject of future work.  

In conclusion, we have mapped the phase diagrams of superfluid $^3$He under nanoscale confinement and have demonstrated the existence of a thermodynamically-stable PDW state, breaking both gauge and translational symmetry.  The stability of both the $A$-phase and the PDW state grow with increasing confinement, consistent with, and serving to reconcile, previous experiments, e.g., the stabilization of the $A$-phase with increasing confinement in Refs.~\cite{Freeman90,Miyawaki00,Kawasaki04}.  Nonetheless, questions remain. What is the nature of the anomalous region at low pressure in the 636 nm device?  Furthermore, as the Helmholtz resonators excite longitudinal (fourth) sound, can we use these to find spectroscopic signatures of the PDW, i.e., order-parameter collective modes \cite{Davis06}?  Finally, our Helmholtz resonators may be capable of observing evidence of Majorana fermions associated with  Andreev bound states at surfaces \cite{Volovik09,Chung09,Wu13,Davis08,Murakawa11} and at the domain walls that make up the PDW observed here, through the superfluid fraction \cite{Wu13} or collective modes \cite{Park15,Mizushima18}---a tantalizing possibility for future experiments.

\begin{acknowledgments}
Authors acknowledge useful conversations with J.A. Sauls, thank W.P. Halperin and Y. Lee for generous assistance with melting curve thermometry, and thank E. Varga for his flow simulation.  This work was supported by the University of Alberta, Faculty of Science and Theoretical Physics Institute; the Natural Sciences and Engineering Research Council, Canada (Grants Nos.~RGPIN-04523-16, RGPIN-2014-4608, DAS-492947-16, and CREATE-495446-17); the Canada Foundation for Innovation; the Canada Research Chair Program; and the Canadian Institute for Advanced Research. A.S. and V.V. contributed equally to this work.
\end{acknowledgments}

%\clearpage
%\pagebreak
%\onecolumn

\renewcommand{\figurename}{Fig.~SM}
\setcounter{figure}{0}

\section*{Appendix A: Low-temperature setup}
 Our low-temperature setup consists of an adiabatic nuclear demagnetization stage, incorporated into a commercial dilution fridge.  The nuclear stage was made from high-purity copper, and vacuum annealed after machining.  The copper stage is attached to the microkelvin plate, where the experimental cell and melting curve thermometer are mounted, via a superconducting heat switch made of high-purity indium wires.  The sample cell housing the liquid $^3$He and Helmholtz resonators is designed after Ref.~\cite{Pollanen12}, with sintered copper powder heat exchangers \cite{Krusius78} of total surface area 84 m$^2$. With $^4$He impurities of 1\% in our $^3$He, this results in a surface coverage of $\approx1/4$ of a monolayer, consistent with diffuse boundary scattering \cite{Freeman90} and the results of our Ginzburg-Landau calculations in the main text.

\section*{Appendix B: Device geometry}

The Helmholtz resonator geometry is depicted in Figure SM 1. The basin diameter, $\diameter$, channel width, $w$, and channel length (after bonding), $l$, are common to all three devices.  The devices differ by their etch depth, and hence helium confinement (see Table \ref{tab:DeviceMeasurements}).

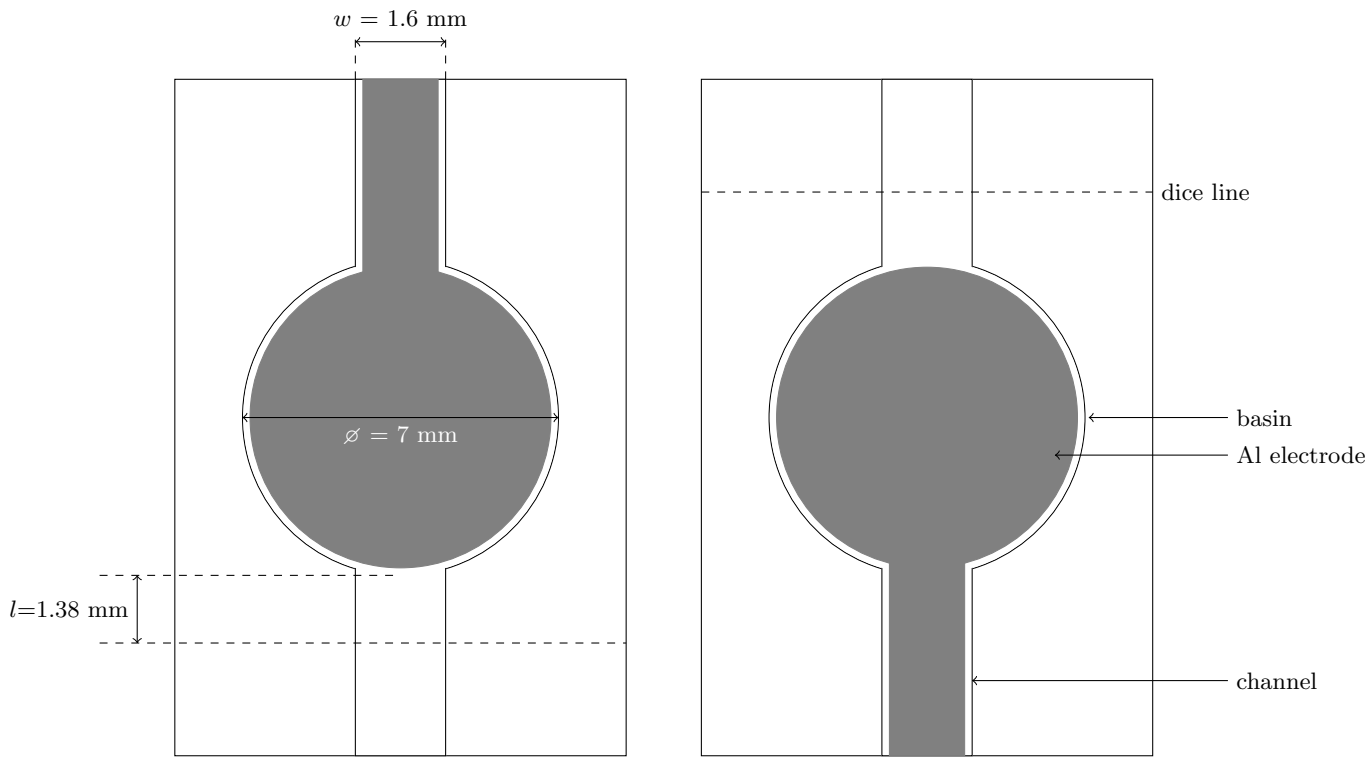
\begin{figure*}[b]
\centering
\begin{tikzpicture}

\begin{scope}[]
\draw (0,0) rectangle (6,9);
\draw (3,4.5) circle (2.1);
\draw[fill=white,white] (2.4,0) rectangle (3.6,9);
\draw (2.4,0) rectangle (3.6,9);
\draw[fill=white,white] (3,4.5) circle (2.082);
\draw[fill=gray,gray] (2.5,4.5) rectangle (3.5,9);
\draw[fill=gray,gray] (3,4.5) circle (2.0);
\draw[dashed] (-1,1.5) -- (6,1.5);
\draw[dashed] (-1,2.4) -- (3,2.4);
\draw[<->] (-0.5,1.5) -- (-0.5,2.4);
\node[left] at (-0.5,1.95) {$l$=1.38 mm};

\draw[<->] (0.9,4.5) -- (5.1,4.5);
\node[below,white] at (3,4.5) {$\diameter$ = 7 mm};

\draw[dashed] (2.4,9) -- (2.4,9.6);
\draw[dashed] (3.6,9) -- (3.6,9.6);
\draw[<->] (2.4,9.5) -- (3.6,9.5);
\node[above] at (3,9.6) {$w$ = 1.6 mm};
\end{scope}

\begin{scope}[shift={(7,0)}]
\draw (0,0) rectangle (6,9);

\draw (3,4.5) circle (2.1);
\draw[fill=white,white] (2.4,0) rectangle (3.6,9);
\draw (2.4,0) rectangle (3.6,9);
\draw[fill=white,white] (3,4.5) circle (2.082);
\draw[fill=gray,gray] (2.5,0) rectangle (3.5,4.5);
\draw[fill=gray,gray] (3,4.5) circle (2.0);

\draw[dashed] (0,7.5) -- (6,7.5) node[right]{dice line};
\draw [<-] (5.15,4.5) -- (7,4.5) node[right]{basin};
\draw [<-] (4.7,4) -- (7,4) node[right]{Al electrode};

\draw [<-] (3.6,1) -- (7,1) node[right]{channel};
\end{scope}
\end{tikzpicture}
\label{fig:ChipGeometry}
\caption{Diagram of a Helmholtz resonator top and bottom chip. The dice line indicates where the chip has been diced to expose the electrode for wiring.}
\end{figure*}

\begin{table*}[]
\centering
\begin{tabular}{|c|c|c|c|c|c|c|c|c|c|}
\hline
\begin{tabular}[c]{@{}c@{}}Device \\ Number\end{tabular} & \begin{tabular}[c]{@{}c@{}}Basin\\ Diameter\\ (mm)\end{tabular} &
\begin{tabular}[c]{@{}c@{}}Electrode\\ Diameter\\ (mm)\end{tabular} &
\begin{tabular}[c]{@{}c@{}}Channel\\ Width\\ (mm)\end{tabular} &
\begin{tabular}[c]{@{}c@{}}Electrode\\ Width\\ (mm)\end{tabular} &
\begin{tabular}[c]{@{}c@{}}Channel\\ Length\\ (mm)\end{tabular} & \begin{tabular}[c]{@{}c@{}}Electrode\\ Height\\ (nm)\end{tabular} & \begin{tabular}[c]{@{}c@{}}Etch\\ Depth\\ (nm)\end{tabular} & \begin{tabular}[c]{@{}c@{}}Basin\\ Confinement\\ (nm)\end{tabular} & \begin{tabular}[c]{@{}c@{}}Channel\\ Confinement\\ (nm)\end{tabular} \\ \hline
1  & 7.00 & 6.80 & 1.60 & 1.40 & 1.38 & 59 & 348 & 577 & 636  \\\hline
2  & 7.00 & 6.90 & 1.60 & 1.50 & 1.38 & 61 & 433 & 745 & 805  \\\hline
3  & 7.00 & 6.80 & 1.60 & 1.40 & 1.38 & 60 & 564 & 1007 & 1067\\\hline
\end{tabular}
\caption{Summary of device geometries.}
\label{tab:DeviceMeasurements}
\end{table*}

It should be noted that the overlap of the two electrodes inside of the basin means that the confinement is $\sim 60$ nm smaller in the basin than in the channel.  While this would result in different behavior for the fluid in the basin, it is irrelevant to our measurement as our experiment is insensitive to the fluid inside the basin.  This is because there is essentially no superflow inside the basin. In Fig.~SM~2, we show a finite element simulation of a Helmholtz resonator velocity profile.  The Helmholtz mode is dominated by the motion of fluid in the channels.  As our measurement probes the superfluid fraction via superflow, it is also dominated by the superfluid in the channels.  Furthermore, since the geometric factor connecting the resonant frequency to the superfluid fraction depends on the area of the Helmholtz resonator in which superflow occurs, any signal due to the small channel-basin interface can be safely discounted. Similarly, the small gap between the electrode and channel wall can be ignored due to it making up a small fraction of the total area.

\begin{figure*}[h]
    \includegraphics[width=.7\textwidth]{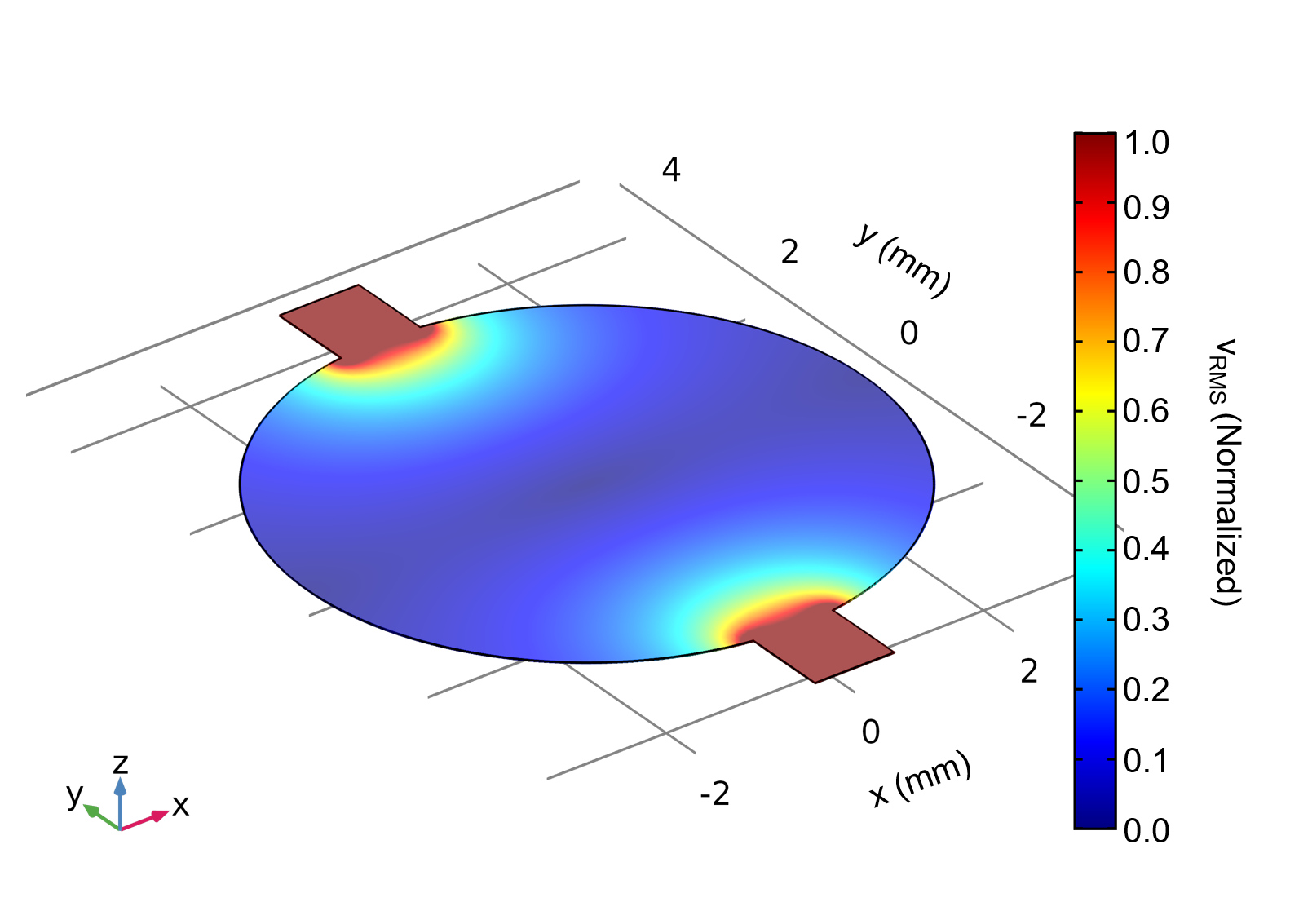}
    \caption{Finite element model of normalized fluid velocity for a Helmholtz resonance.  The Helmholtz resonance is dominated by motion of the fluid in the channel, hence the resonance frequency in our experiment is dominated by the superflow in the channel.  }
 \label{fig:flow}
 \centering
\end{figure*}

\section*{Appendix C: Thermal Effects}
A detailed model of thermal effects in our Helmholtz resonator geometry has been developed in Ref.~\citenum{Souris17}.  This model concerns the thermal resistance between the helium in the basin and the surrounding helium via the helium in the channel, as well as the thermal resistance at the boundary between the helium and solid surfaces (called the Kapitza resistance \cite{Kapitza41}) and the thermal resistance through the quartz.  Heat flux through the chip meets boundary resistance at each of the material interfaces: $^3$He to aluminum, aluminum to quartz (which is negligibly small), and quartz to $^3$He. Additionally there is some thermal resistance while propagating through each material.

To estimate the Kapitza resistance, Bekarevich and Khalatnikov have developed a model for thermal resistance between $^3$He and a solid surface that is applicable for temperatures below 0.2 K where energy flux is transmitted primarily through collective excitations of zero sound \cite{Bekarevich61}. The model predicts the resistance over an area $A_{\text{tot}}$ to be
\begin{equation}
    A_{\text{tot}}R_KT^3 = \frac{5h^3}{8\pi^5k_B^4} \left(\frac{\rho}{\rho_{\text{He}}} \right) \frac{c^3}{v_F} \left(\frac{10^{-7}}{aF+b\Phi}\right),
    \label{eq:KapitzaModel}
\end{equation}
where $\rho$ and $c$ are the density and transverse speed of sound in the solid (i.e. aluminum or quartz), $v_F$ is the Fermi velocity of $^3$He, $a$ and $b$ are constants describing the diffusivity of quasiparticle scattering, and $F$ and $\Phi$ are material constants which depend on the ratio of the transverse and longitudinal speeds of sound in the solid. Following the analysis of copper by Anderson et al.~\cite{Anderson64}, we take $F=1.6$, $\Phi=1$, $a=0.38$, and $b=0.05$. The difference in these values for aluminum and quartz is expected to be relatively small. 

We consider the resulting Kaptiza resistance at 20 bar, where we measure the largest gap between the two observed phase transitions. At this pressure $\rho_{\text{He}} = 108.7$ kg/m$^3$ and $v_F = 38.7$ m/s. The material parameters for aluminum are taken to be $\rho_{\text{Al}} = 2730$ kg/m$^3$ and $c_{\text{Al}} = 3240$ m/s. For quartz, they are $\rho_{\text{Q}} = 2650$ kg/m$^3$ and $c_{\text{Q}} = 2210$ m/s. The total area of one side of the device including the basin and two channels is $4.29\times 10^{-5}$ m$^2$. At $1.6$ mK, approximately where the PDW to B-phase transition occurs, the thermal boundary resistance for aluminum and quartz respectively are thus
\begin{equation}
    R_{K,\text{Al}} = 3.12 \times 10^{5}~\textrm{K/W}
\end{equation}
and
\begin{equation}
    R_{K,\text{Q}} = 9.62 \times 10^{4}~\textrm{K/W}.
\end{equation}
Each of these boundary resistances is added in series with the thermal resistances of aluminum, $R_{\text{Al}}$, and quartz, $R_{\text{Q}}$, to get the chip resistance. Given that the electrode thickness is four orders of magnitude smaller than the quartz thickness it can safely be neglected. Taking into account the two chip surfaces and two superfluid channels, which are all added in parallel, we find the total thermal resistance to be
\begin{equation}
    R_{\text{tot}} = \left( \frac{2}{R_{K,\text{Al}}+R_{\text{Q}}+R_{K,\text{Q}} } +  \frac{2}{R_{\text{He}}} \right)^{-1}.
\end{equation}
The thermal resistance across $0.5$ mm of quartz is found to be $R_{\text{Q}} = 758$ K/W by extrapolating data from \cite{Vollhardt13} to $1.6$ mK. For superfluid $^3$He-B at 20 bar the thermal resistance is obtained from \cite{Hall86}. The result is found to be $R_{\text{He}}= 2.07\times 10^{4}$ K/W. Adding these resistances according to Eq.~4 gives
\begin{equation}
    R_{\text{tot}} = 9.85 \times 10^3~\textrm{K/W}.
\end{equation}
Therefore, the power required to produce a thermal gradient equal to the temperature gap between transitions of $0.2$ mK would be 
\begin{equation}
    \dot{Q} = \frac{\Delta T}{R_{\text{tot}}} = 20.3~\textrm{nW}.
\end{equation}
This number rules out the possibility of heating effects contributing to our signal for the following reasons.  First, this number would have to apply to each device, since they each have identical electrodes, and would imply that the heat leak from the measurement alone would have to be larger than 60.9 nW.  This is roughly 200 times larger than expected heat leak into the entire nuclear demagnetization stage \cite{Haard01}. Furthermore, prior experimental evidence \cite{Harrison79} suggests that Eq.~\ref{eq:KapitzaModel} typically over estimates the thermal resistance, which means the actual required power would be even higher than predicted. This calculation should therefore be taken as a lower limit on the power required to produce the observed temperature gap.

Second, such heating could not explain the different widths of the PDW state shown in the main text.  Since the electrodes in each device are identical, one expects each device to have identical heating and hence produce identical thermal gradients, inconsistent with the varying widths of the PDW phase observed in Fig.~4 of the main text.  Third, in our geometry heating effects would result in a thermal gradient across the helium channel, which is the only helium that contributes to the Helmholtz signal, and not two distinct regions of temperature.  Hence from such a heating model, one would expect a single broad phase transition, and not two sharp phase transitions.  Fourth, one cannot reconcile this level of heating from Joule heating as our electrodes are superconducting inside the basin and along the channel.  The only Ohmic losses in our electrical circuit would exist where the measurement leads are silver epoxied to the electrodes, but this contact is made in the surrounding bath of helium and cannot produce heating inside the helium channel of the Helmholtz resonator.  Fifth and finally, we note that the temperatures at which the observed phase transitions occur do not depend on the measurement drive, further ruling out any thermal effects from the measurement.

In light of these thermal calculations and experimental observations, we conclude that the stable phase transitions observed in our devices are genuine and do not result from any untoward thermal effects. 

\end{document}